\documentclass[conference]{IEEEtran}
\ifCLASSINFOpdf
\else
\fi

\usepackage{graphicx} 
\usepackage{pdfpages}
\usepackage{subcaption}
\graphicspath{ {images/} }
\usepackage{float} 

\begin{document}
\title{Evaluation of Suitability of Different\\ Transient Stability Indices for Identification of\\ Critical System States}

\author{\IEEEauthorblockN{A. Sajadi, \textit{Member, IEEE},  R. Preece, \textit{Member, IEEE}, and J. V. Milanovi\'{c}, \textit{Fellow, IEEE}}
\IEEEauthorblockA{School of Electrical and Electronic Engineering\\
University of Manchester\\
Oxford Rd, Manchester M13 9PL, UK\\
Corresponding Author: sajadi@manchester.ac.uk}}

\maketitle

\begin{abstract}
Power system stability indices are used as measures to evaluate and quantify the response of the system to external large disturbances. This paper provides a comparative analysis of established transient stability indices. The indices studied in this paper include rotor-angle difference based transient stability index (TSI), rate of machine acceleration (ROMA), transient kinetic energy (TKE), and transient potential energy (TPE). The analysis is performed using the 3-machine, 9-bus standard test system under a realistic range of loading levels. The aim of the study is to determine their suitability for reliable identification  of critical system conditions considering system uncertainties.
\end{abstract}


\IEEEpeerreviewmaketitle

\section{Introduction}


Transient stability analysis investigates the dynamic behavior of a given system in respect to the time following a large external disturbance. The external disturbance could be in form a short term fault, such as short circuit faults on transmission lines or generators with a successful clearance, or a long term fault, such as an outage of generation unit(s) or a disconnection of lines \cite{machowski2011power}. Following a fault, regardless of its type, the oscillations that are excited by the fault should be damped such that ringing decays within the first few cycles to few seconds following the fault. Otherwise, the transient behavior of the system may dominate the system response that the system trajectory diverges from the stability region associated with the pre-fault equilibrium point and potentially lead to system- wide failures  \cite{sajadi2016transient}. Therefore, transient stability analysis are a key stage in planning and operation studies of power systems.

Transient stability index is a measure to quantify the distance between any given operating point of the system (pre-fault equilibrium point) and the critical operating point of the system (the margin of the stability region). In other words, it is an indication of power system stability limit at any given operating point. Thus, the accuracy of this measure is crucially important for power system industry to ensure a reliable, stable and secure delivery of power to  consumers.


Future power systems will be associated with a greater degree of uncertainty and complexity because of involvement of highly intermittent power generation sources and power- electronics and overall change in the paradigm  of power networks. The higher complexity of the system can be also interpolated as higher dimensionality of the system. As a result, it is essential to develop new probabilistic analytical tools to quantify involved risk in future power systems operation and ensure stability and security of electrical energy delivery.


The goal of this work is to identify and validate robust stability indicator(s) that can be utilized for different facets of power system stability risk analysis. There is significant interest in identifying critical system conditions and understanding the corrective action required to reduce their criticality. A proper understanding of the performance of different transient stability indicators will enable fast and reliable identification of appropriate actions.


To reach this goal, transient stability indices introduced and established in literature are compared. The considered indices include rotor-angle difference based transient stability index (TSI) \cite{Shi09p1}, rate of machine acceleration (ROMA) \cite{teleginaimpact}, transient kinetic energy (TKE) \cite{kundur1994power}, and transient potential energy (TPE) \cite{saunders2014transient}. To ensure greater applicability, only simulation-based methods have been investigated as they are more easily implemented on existing practical network models and will inherently include the effects of discontinuous system elements (such as controller limits or saturation effects). The analytical methods, such as the use of energy functions, are not included in present analysis due to their computational complexity and limited applicability for large and complex power systems.

Numerical results are illustrated on a 3-machine, 9-bus standard test system \cite{9busonline} under a range of feasible loading levels. 
The disturbances considered in this study include power system faults at various busbars representing various levels of event's severity. 
Then, corresponding transient stability indices are computed for all of the studied loading levels and considered faults in all dimensions of operational space. Finally, the $sensitivity$ and $smoothness$ of each index is measured to evaluate their suitability for further probabilistic studies. In further work, the most suitable metric will be used as indicator to predict a broader probabilistic surface and region of stability for system operation. This will be helpful to identify the risks and costs associated with operation of the system.


The contribution of this study is to identify the most practical and efficient transient stability index to conduct further studies regarding operation and control of power systems with a higher degree of complexity and uncertainty. In particular, the aim is to identify a metric that accurately reflects the transient stability in terms of sensitivity to system parameter changes in higher dimensions and fault severity. This study considers change of loading level as the only variable change.



\section{Transient Stability Indices}


%

\begin{figure}[h]
	\centering
	\includegraphics[width=0.45\textwidth]{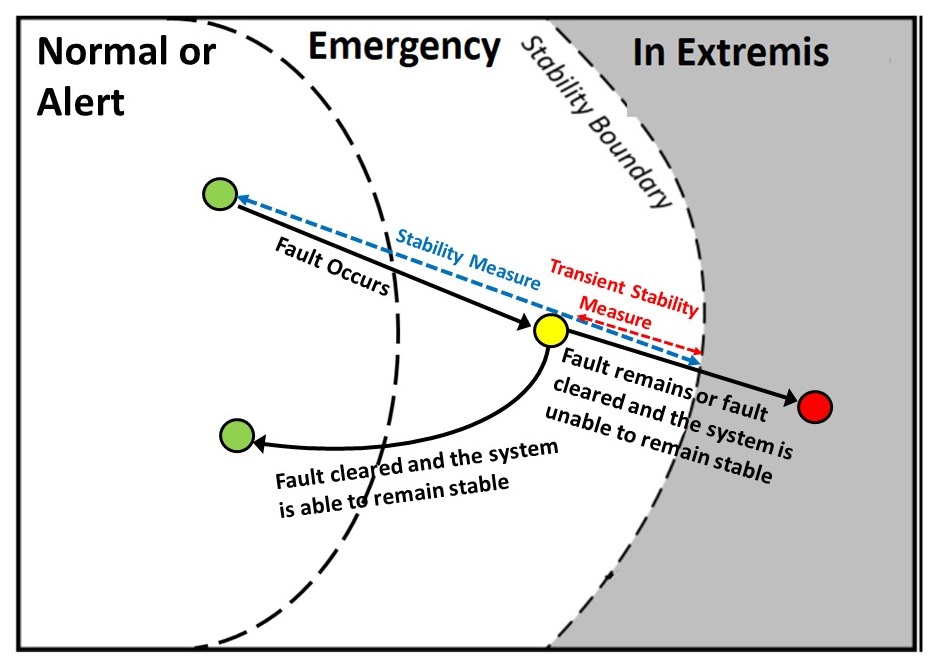}
	\caption{Conceptual visualization of transient stability indices in power systems}
	\label{visualization}
\end{figure}

Transient stability can be defined by the system's ability to maintain its operation following a fault \cite{machowski2011power}. 
And the longest time that system's trajectory remains within the stability region associated with the pre-fault equilibrium point before reaching the critical operating point at which the instability begins, is called critical clearing time (CCT) \cite{machowski2011power}. Considering these definitions, the transient stability index can be defined by a quantification of system's strength to sustain its transient stability. Fig. \ref{visualization} illustrates a visualization of transient stability index.

The following section briefly describes the time-domain based transient stability indices considered in this comparative study. The full theoretical background on these indices are extensively described in the quoted references.

\subsection{Rotor Angle Difference Based Transient Stability Index (TSI)}

This index relies on maximum rotor angle separation between any two given generators and is given by (\ref{eq:TSI}) \cite{Shi09p1}.

\begin{equation} \label{eq:TSI}
TSI =\frac{360-\delta_{max}}{360+\delta_{max}} \times 100
\end{equation}

In (\ref{eq:TSI}), $\delta_{max}$ is maximum rotor angle difference between any two generators in the system immediately after fault inception. The closer the value of $TSI$ to 100 is, the more stable the power system is. 

\subsection{Rate of Machines Acceleration (ROMA)}

The rate of acceleration or deceleration of generators' rotors in a power system is an indication of its inertia and, therefore, the rate of the frequency deviation \cite{teleginaimpact}. Thus, the rate of machines' acceleration, similar to $Rate~of~Change~of~Frequency~(ROCOF)$ \cite{ROCOFreport}, can be defined by (\ref{eq:ROCOF}).

\begin{equation} \label{eq:ROCOF}
ROMA =max \Bigg \langle \frac{da_{PFT_i}}{dt} \Bigg \rangle  \approx max \Bigg \langle \frac{\Delta a_{PFT_i}}{\Delta t} \Bigg \rangle 
\end{equation}

In (\ref{eq:ROCOF}), $\Delta a_{PFT_i}$ and $\Delta t$ are finite differences of rotor acceleration and time of $i$-th machine immediately after fault occurrence to approximate the differentials.

\subsection{Transient Kinetic Energy (TKE)}

The generators' transient kinetic energy immediately after the fault clearance is defined by (\ref{eq:TKE}) \cite{kundur1994power}. 

\begin{equation} \label{eq:TKE}
TKE =  \sum_{i=1} \frac{1}{2} ~ J_i \cdot \Delta\omega_i^2
\end{equation}

In (\ref{eq:TKE}), $J_i$ and $\Delta\omega_i$ are angular momentum of the rotor at synchronous speed and speed deviation of $i$-th generator. 

\subsection{Transient Potential Energy (TPE)}

The generators' transient potential energy immediately after the fault clearance is defined by (\ref{eq:TPE}) \cite{saunders2014transient}.

\begin{equation} \label{eq:TPE}
TPE =\int_{t_{fault}}^{t_{clear}} \Big[\Delta P_{G_i} - \Delta P_{G_j} \Big] \Delta f_{ij} \cdot dt
\end{equation}

In (\ref{eq:TPE}), $\Delta P_{G_i}$ and $\Delta P_{G_j}$ refer to transient active power of any given pair of generators $i$ and $j$ and $\Delta f_{ij}$ is the frequency difference between them. $t_{fault}$ and $t_{clear}$ is the time at which fault occurs and clears, respectively.

\section{Computational Implementation}

In this section, the case study used in this paper and considerations and data toolkit used for computational implementation are described.

\subsection{Power Systems Simulation}

\begin{figure}[h]
	\centering
	\includegraphics[width=0.28\textwidth]{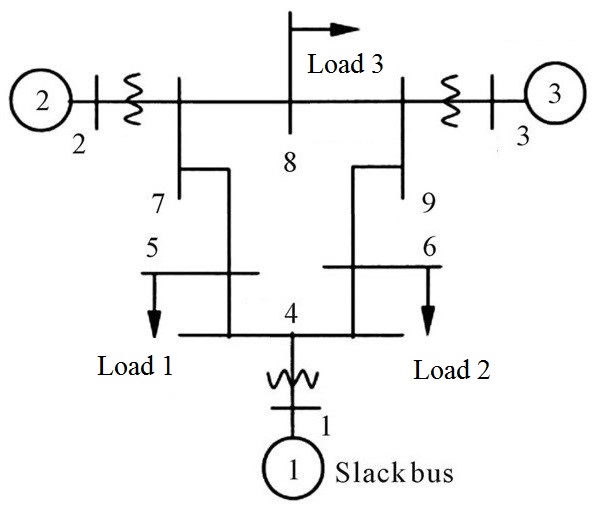}
	\caption{One-line diagram of a 3-machine, 9-bus standard test system}
	\label{9bus_diagram}
\end{figure}

This study used a 3-machine, 9-bus standard test system (also known as P. M. Anderson 9-Bus), shown in Fig. \ref{9bus_diagram}, as a case study. This system consists of 3 synchronous machines and 3 loads \cite{Andersonsystem}.

The power system simulation was carried out in DigSILENT PowerFactory software package, v15.2.8. Based on previously carried out studies and available information in literature \cite{chiang1995theoretical}, \cite{mariotto2010power}, \cite{dhole2005antigen}, five faults were considered, as described in Table \ref{table:faults}.

\begin{table}[h]
	\centering
	\caption{Description of Faults }
	\label{table:faults}
	\begin{tabular}{ccc}
		\hline 
		Fault No. & Faulted Bus & Faulted Line \\ \hline
		1         & 4           & 4-6          \\\hline
		2         & 5           & 5-7          \\\hline
		3         & 6           & 6-9          \\\hline
		4         & 7           & 7-8          \\\hline
		5         & 8           & 8-9          \\\hline
	\end{tabular}
\end{table}

The applied faults were balanced 3-phase faults with clearance time of 10 cycles.

The load and generation dispatch datasets were generated using optimal power flow (OPF) in MATPOWER package. The load and generation data were developed for three scenarios: 

\begin{itemize}{ 
	\item \textbf{1-dimensional surface:} By proving 1 degree of freedom in change of loads, corresponding to change of one load as the only variable while two other loads remain fixed.
	
	\item \textbf{2-dimensional surface:} By proving 2 degrees of freedom in change of loads, corresponding to change of two loads as the variables while the other load remains fixed.
	\item \textbf{3-dimensional surface:} By proving 3 degrees of freedom in change of loads, corresponding to change of all three loads as the variables.}
\end{itemize}

To generate load and generation datasets, for each scenario, loads were varied from  30\% to 100\% of their nominal consumption capacity with steps of 2\%, resulting in investigation of 46,656 operating conditions for each fault. For each step, an OPF solution with a homogeneous cost function for all generators was run. The reason for using 2\% steps was to have a reasonable computational process time with sufficient data point to capture the continuous possible load variation.

Finally, the datasets were used to run the electromagnetic transient (EMT) simulation in DigSILENT PowerFactory. 

\subsection{Data Analytics}

The overarching aim of this study is to identify the transient stability indices that have the greatest potential for further use when identifying critical system conditions. These indices must therefore be sensitive to parameter changes and vary as conditions vary. Moreover, they must also be smooth in these variations in order to provide confidence that they are useful as predictive indices. This is of particular importance when the study expands to a multi-dimensional search (in multiple parameters) and a multi-dimensional surface (and not only a line) is produced. 
In this way, it is more likely that the global, rather than local minima (with respect to stability performance) is identified.  

To compute the transient stability indices, discussed in previous section, and measure their features, the results from the EMT simulation in DigSILENT PowerFactory simulation tool were imported in MATLAB and aforementioned transient stability indices were computed. 

Following section describes the toolkits used for data analytics after computation of these indices.

\subsubsection{Data Standardization}

Since this is a comparative study, the computed transient stability indices are required to be standardized as they may vary in different ranges. Data normalization refers to adjusting data measured on different scales to a common scale to bring them into a meaningful alignment. The standardization technique sued in this study is given by (\ref{eq:standardize}).

\begin{equation} \label{eq:standardize}
x' =  \frac{x}{max(x)}
\end{equation}

In (\ref{eq:standardize}), $x'$ and $x$ are standardized data and original data points, respectively, and $max(x)$ is maximum value of the dataset that $x$ belongs to. 

{It should be emphasized that the data standardization technique used in this study is to only cap the datasets at their maximum value as a common reference point. A rescale (normalization) between maximum and minimum will mask of the actual sensitivity and smoothness of the datasets. Standardizing at the maximum value avoids this.}

\subsubsection{Sensitivity Analysis}

To measure the sensitivity of each output dataset $y$ to a change of operational variables, load in this study denoted $x$, sensitivity index (SNI) defined by (\ref{eq:SNI}) was used \cite{Hooker_data}.

\begin{equation} \label{eq:SNI}
SNI =- \log_{10} \Bigg \langle \int \frac{dy}{dx} ~ dx \Bigg \rangle
\end{equation}

In (\ref{eq:SNI}), $\frac{dy}{dx}$ is the instantaneous slope of $y(x)$. And integration is to measure the size of the slope for all data points. The smaller the value of $SNI$ is, the more sensitive the dataset is to a change of variables. { The logarithmic scale is to avoid appearance of a significant order of decimal digits in the results. It should be noted that this is a comparative study and the scale used for comparison of the results does not influence the outcome of this research.}

In a $n$-dimensional search space, the total sensitivity of the overall surface is given as (\ref{eq:SNIn}).

\begin{equation} \label{eq:SNIn}
SNI_{overall} =\sum_{i=1}^{n} \frac{SNI_i}{n!}
\end{equation}

In (\ref{eq:SNIn}), $SNI_i$ is the $SNI$ of the dataset of $n$-th dimension.

\subsubsection{Smoothness Analysis}

To measure the smoothness of each output dataset $y$ to a change of operational variables, $x$ (load in this study), a smoothness index (SMI) defined by (\ref{eq:SMI}) was used \cite{Hooker_data}.

\begin{equation} \label{eq:SMI}
SMI =- \log_{10} \Bigg \langle \int \bigg[\frac{d^2y}{dx^2} \bigg]^2 dx \Bigg \rangle
\end{equation}

In (\ref{eq:SMI}), $\frac{d^2y}{dx^2}$ is the curvature of $y(x)$. And integration is to measure the size of the curvature for all data points. The greater the value of $SMI$ is, the smoother the dataset is. {The logarithmic scale, similar to the $SNI$, is to avoid appearance of a significant order of decimal digits in the results.}

In a $n$-dimensional search space, the total smoothness of the overall surface is given as (\ref{eq:SMIn}).

\begin{equation} \label{eq:SMIn}
SMI_{overall} =\sum_{i=1}^{n} \frac{SMI_i}{n!}
\end{equation}

In (\ref{eq:SMIn}), $SMI_i$ is the $SMI$ of the dataset of $n$-th dimension.

\section{Results and Discussions}

In this section, the results obtained using different transient stability indices for assessment of transient stability of the 3-machine, 9-bus standard test system shown in Fig. \ref{9bus_diagram}, are presented.

%
%
%
%

\subsection{Data Standardization}

\begin{figure}[h]
	\centering
	\includegraphics[width=0.5\textwidth]{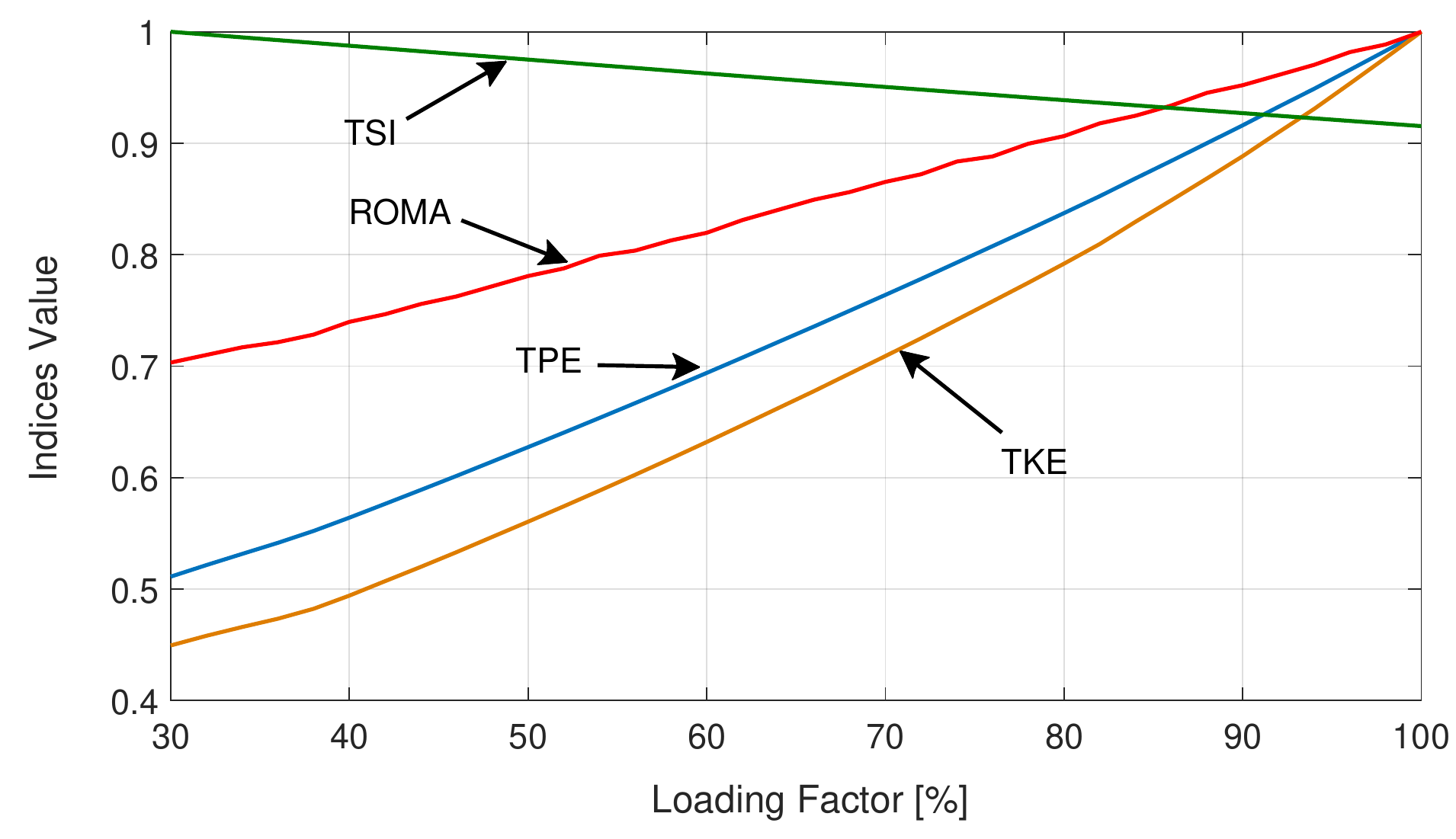}
	\caption{Standardized values of studied transient stability indices in a single-dimensional analysis: Load 1 changes as the only variable, following fault 1 in the studied test system}
	\label{standardized}
\end{figure}

In this study, the data from investigated transient stability indices were standardize at the 100\% loading level of system, for all three loads, as the common reference point. Fig. \ref{standardized} illustrates the standardized values of studied indices in one-dimension with load 1 as the only variable, following fault 1.

{The plots illustrated in Fig. \ref{standardized} represent studied stability indices. It is evident that the indices were standardized with a common reference point of 1 while their range is mapped relatively. Amongst the indices, $TSI$ is the only index whose value increases as the loading level of the system decreases. Whereas the values of $ROMA$, $TKE$, and $TPE$ decreases proportional to decrease of loading level of system.}
	
In a physical sense, the lower the loading level of the system is, the lower amount of transient energy in the network to dissipate following the fault clearance is. As a result, transient kinetic and potential energies and machines' acceleration will be reduced proportional to lower levels of system loading. Consequently, the rotor angle differences in the system will reduce leading to an increase of $TSI$ value. 

{Thus, the lighter the loading level of system becomes, the less severe the faults become as the transient stability margin of system increases. 

The results from analytical investigation and multi-dimensional analysis on these indices are shown in next two sections.

\subsection{Sensitivity Analysis} \label{sec:sens}

In this section, results from sensitivity analysis of the studied stability indices are presented. As a reminder, the smaller the value of $SNI$ is, the more sensitive the dataset is to a change of variables. {Sensitivity of a transient stability index is crucially important as it ensures an accurate reflection of the transient stability in terms of sensitivity to system parameter changes.}

\subsubsection{One-dimensional Analysis} 

{The values of computed $SNI$ for single-dimensional surface corresponding to a change of load 1 as the only variable are presented in Table \ref{table:1D_sens}. It should be noted that results from a change of load 2 and load 3 as the only the variable in the system with other two loads kept constant, are similar.}

\begin{table}[h]
	\centering
	\caption{The results from sensitivity analysis of stability indices -- single dimensional -- {Change of load 1 as the only variable while loads 2 and 3 remain constant}}
	\label{table:1D_sens}
	\begin{tabular}{ccccc} \hline
		
\multicolumn{1}{|c|} {Index}         &\multicolumn{1}{c|}  {TSI}      & \multicolumn{1}{c|} {ROMA}      &\multicolumn{1}{c|}  {TKE}    &\multicolumn{1}{c|}  {TPE} \\ \hline

\multicolumn{1}{|c|}{Fault 1} &	\multicolumn{1}{c|} {$2.59$}    &\multicolumn{1}{c|} {$1.94$}    &\multicolumn{1}{c|} {$1.63$}    &\multicolumn{1}{c|} {$1.76$}    \\\hline
\multicolumn{1}{|c|}{Fault 2} &	\multicolumn{1}{c|} {$2.61$}    &\multicolumn{1}{c|} {$1.85$}    &\multicolumn{1}{c|} {$1.55$}    &\multicolumn{1}{c|} {$1.68$}    \\\hline
\multicolumn{1}{|c|}{Fault 3} &	\multicolumn{1}{c|} {$2.59$}    &\multicolumn{1}{c|} {$1.97$}    &\multicolumn{1}{c|} {$1.54$}    &\multicolumn{1}{c|} {$1.74$}    \\\hline
\multicolumn{1}{|c|}{Fault 4} &	\multicolumn{1}{c|} {$2.49$}    &\multicolumn{1}{c|} {$2.11$}    &\multicolumn{1}{c|} {$1.59$}    &\multicolumn{1}{c|} {$1.75$}    \\\hline
\multicolumn{1}{|c|}{Fault 5} &	\multicolumn{1}{c|} {$2.47$}    &\multicolumn{1}{c|} {$2.25$}    &\multicolumn{1}{c|} {$1.60$}    &\multicolumn{1}{c|} {$1.78$}    \\\hline
		
	\end{tabular}
\end{table}

From the results shown in Table \ref{table:1D_sens}, it can be seen that $TKE$, consistently, holds the minimum $SNI$ values amongst the studied indices for all studied faults, ranging from 1.54 to 1.63. Whereas the $TSI$ indicates the highest $SNI$ values, consistently, for all faults, ranging from 2.47 to 2.61, outlining 60\% greater values of $SNI$ than $TKE$ does. The $SNI$ values for $TPE$ and $ROMA$ are second and third highest. 

It can be concluded that the in single-dimensional analysis, $TKE$ is the most sensitive index and $TPE$, $ROMA$, and $TSI$ follow.

\subsubsection{Two-dimensional Analysis}

{The values of computed $SNI$ for two-dimensional surface corresponding to a change of loads 2 and 3 as the only variables are presented in Table \ref{table:2D_sens}. It should be noted that results from a change of loads 1 and 3 and loads 1 and 2 as the only the variables in the system with other load kept constant, are similar.} 

\begin{table}[h]
	\centering
	\caption{The results from sensitivity analysis of stability indices -- two dimensional -- {Change of loads 2 and 3 as the variables while load 1 remains constant}}
	\label{table:2D_sens}
	\begin{tabular}{ccccc} \hline
		
\multicolumn{1}{|c|} {Index}         &\multicolumn{1}{c|}  {TSI}      & \multicolumn{1}{c|} {ROMA}      &\multicolumn{1}{c|}  {TKE}    &\multicolumn{1}{c|}  {TPE} \\ \hline

\multicolumn{1}{|c|}{Fault 1}&\multicolumn{1}{c|} {$2.62$}    &\multicolumn{1}{c|} {$2.01$}    &\multicolumn{1}{c|} {$1.81$}    &\multicolumn{1}{c|} {$1.81$}    \\\hline
\multicolumn{1}{|c|}{Fault 2}&\multicolumn{1}{c|} {$2.69$}    &\multicolumn{1}{c|} {$2.01$}    &\multicolumn{1}{c|} {$1.71$}    &\multicolumn{1}{c|} {$1.80$}    \\\hline
\multicolumn{1}{|c|}{Fault 3}&\multicolumn{1}{c|} {$2.64$}    &\multicolumn{1}{c|} {$2.03$}    &\multicolumn{1}{c|} {$1.68$}    &\multicolumn{1}{c|} {$1.82$}    \\\hline
\multicolumn{1}{|c|}{Fault 4}&\multicolumn{1}{c|} {$2.68$}    &\multicolumn{1}{c|} {$2.18$}    &\multicolumn{1}{c|} {$1.88$}    &\multicolumn{1}{c|} {$1.91$}    \\\hline
\multicolumn{1}{|c|}{Fault 5}&\multicolumn{1}{c|} {$2.65$}    &\multicolumn{1}{c|} {$2.21$}    &\multicolumn{1}{c|} {$1.90$}    &\multicolumn{1}{c|} {$2.01$}    \\\hline

	\end{tabular}
\end{table}


The $SNI$ values for $TKE$, consistently, are the minimum amongst the studied indices for all studied faults, ranging from 1.68 to 1.90. The $SNI$ values for $TSI$ are the highest amongst the studied indices for all faults, ranging from 2.64 to 2.69. The $SNI$ values for $TPE$ and $ROMA$ are second and third highest by ranging within 1.80 and 2.01 and 2.01 and 2.21.

It can be concluded that the in two-dimensional analysis, $TKE$ is the most sensitive index and $TPE$, $ROMA$, and $TSI$ follow. 

\subsubsection{Three-dimensional Analysis}

{The values of computed $SNI$ for sensitivity analysis of a three-dimensional surface corresponding to a change of loads 1, 2 and 3 as the variables are presented in Table \ref{table:3D_sens}. }

\begin{table}[h]
	\centering
	\caption{The results from sensitivity analysis of stability indices -- three dimensional  -- {Change of loads 1, 2 and 3 as the variables}}
	\label{table:3D_sens}
	\begin{tabular}{llllllllllllllllllll} \hline
\multicolumn{1}{|c|} {Index}  & \multicolumn{1}{c|}{TSI} & \multicolumn{1}{c|}{ROMA} & \multicolumn{1}{c|}{TKE} & \multicolumn{1}{c|}{TPE}  \\ \hline 
		
\multicolumn{1}{|c|}{Fault 1}&\multicolumn{1}{c|} {$2.70$}    &\multicolumn{1}{c|} {$2.03$}    &\multicolumn{1}{c|} {$1.86$}    &\multicolumn{1}{c|} {$1.88$}    \\\hline
\multicolumn{1}{|c|}{Fault 2}&\multicolumn{1}{c|} {$2.75$}    &\multicolumn{1}{c|} {$2.02$}    &\multicolumn{1}{c|} {$1.86$}    &\multicolumn{1}{c|} {$1.88$}    \\\hline
\multicolumn{1}{|c|}{Fault 3}&\multicolumn{1}{c|} {$2.71$}    &\multicolumn{1}{c|} {$2.03$}    &\multicolumn{1}{c|} {$1.85$}    &\multicolumn{1}{c|} {$1.89$}    \\\hline
\multicolumn{1}{|c|}{Fault 4}&\multicolumn{1}{c|} {$2.69$}    &\multicolumn{1}{c|} {$2.15$}    &\multicolumn{1}{c|} {$1.91$}    &\multicolumn{1}{c|} {$1.96$}    \\\hline
\multicolumn{1}{|c|}{Fault 5}&\multicolumn{1}{c|} {$2.66$}    &\multicolumn{1}{c|} {$2.09$}    &\multicolumn{1}{c|} {$1.92$}    &\multicolumn{1}{c|} {$1.96$}    \\\hline

	\end{tabular}
\end{table}

The results shown in Table \ref{table:3D_sens} indicate a similar information as was previously indicated in Tables \ref{table:1D_sens} and \ref{table:2D_sens}. $TKE$ and $TSI$ consistently feature the smallest and largest $SNI$ values for all studied faults which reflects the highest and lowest sensitivity amongst studied indices, respectively, by ranging within 1.85 and 1.92 for $TKE$ and 2.66 and 2.75 for $TSI$. $TPE$ and $ROMA$ stand second and third in ranking of sensitivity of indices.

$TKE$ shows the greatest level of sensitivity because it is a function of the generators' angular momentum and the square of the speed deviation. The greater the loading of the system is, the greater speed deviations are. As a result, this index increases with the square of the system loading. Similarly, $TPE$ is a quadratic function of the system loading as this index is computed as the product of frequency deviation and active power deviation of a pair of generators. $ROMA$ is the third sensitive index and is a first order function of the system loading. This index is computed using the generators’ acceleration which a function of system’s loading. Finally, $TSI$ is the least sensitive index as is a first order function of the angular difference of generators. The generators' angles vary proportionally to the system loading with a similar rate and, therefore, the difference between them (and therefore the $TSI$) changes at a slower rate compared to the other studied indices.

\begin{figure}[h]
	\centering
	\includegraphics[width=0.445\textwidth]{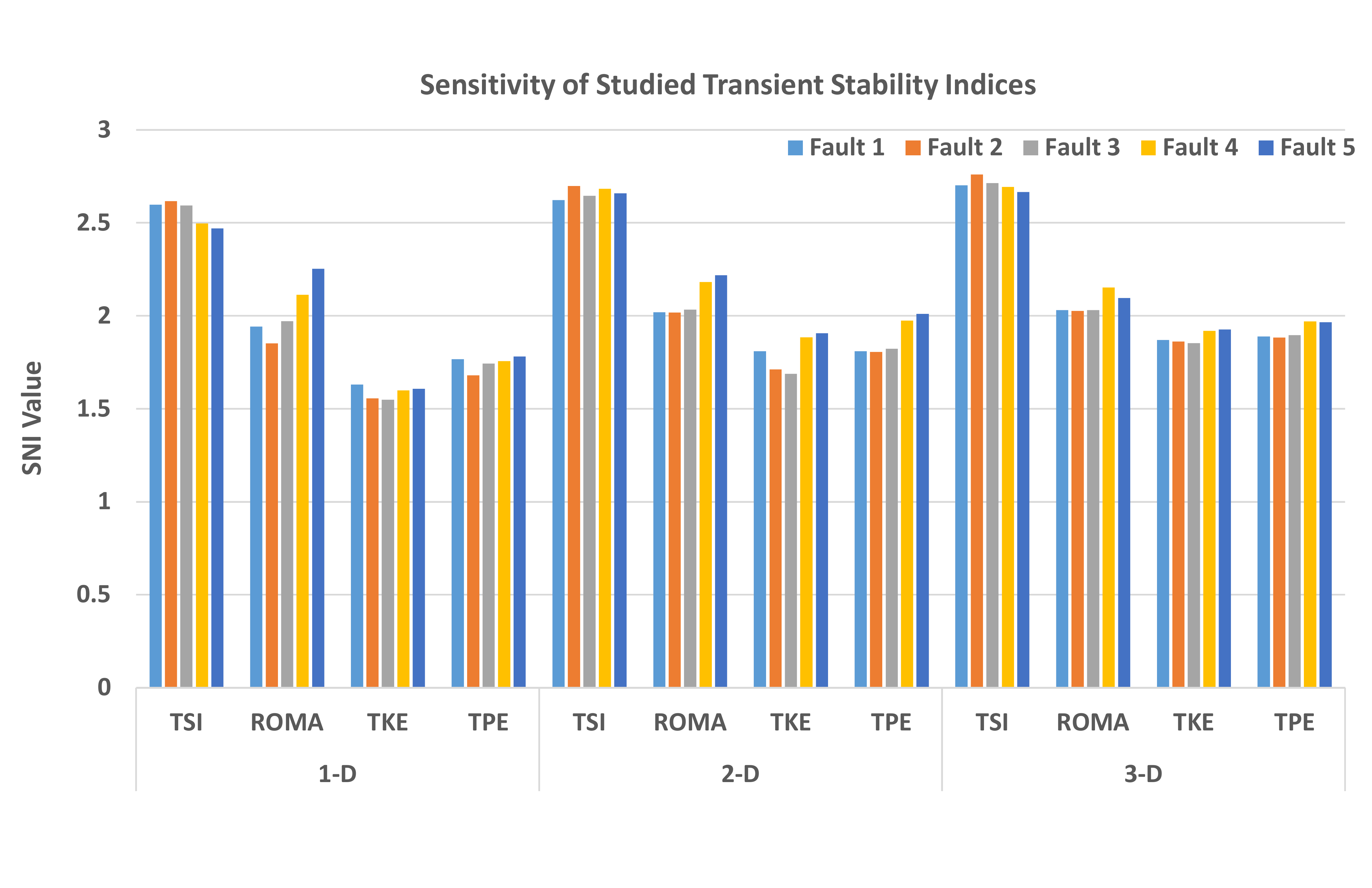}
	\caption{The results from sensitivity analysis of studied transient stability indices}
	\label{Sens_grpah}
\end{figure}

Fig. \ref{Sens_grpah} visualizes the $SNI$ for studied indices in multiple dimensions. The results shown in this figure reveal the consistency of used sensitivity measure and the appropriateness of using this measure for this work as it has been suitably adapted for multi-dimensional analysis.

\subsection{Smoothness Analysis} \label{sec:smooth}

In this section, results from smoothness analysis of the studied stability indices are presented. As a reminder, the greater the value of $SMI$ is, the smoother the dataset is. {Smoothness of a transient stability index is very important as it ensures its suitability for further probabilistic studies of the system as the system parameters change in multiple dimensions.}

\subsubsection{One-dimensional Analysis}

{The values of computed $SMI$ for single-dimensional surface corresponding to a change of load 1 as the only variable are presented in Table \ref{table:1D_smooth}. The results from a change of load 2 and load 3 as the only the variable in the system with other two loads kept constant, are similar. }

\begin{table}[h]
	\centering
	\caption{The results from smoothness analysis of stability indices -- single dimensional -- {Change of load 1 as the only variable while loads 2 and 3 remain constant}}
	\label{table:1D_smooth}
	\begin{tabular}{ccccc} \hline
		
\multicolumn{1}{|c|} {Index}         &\multicolumn{1}{c|}  {TSI}      & \multicolumn{1}{c|} {ROMA}      &\multicolumn{1}{c|}  {TKE}    &\multicolumn{1}{c|}  {TPE} \\ \hline

\multicolumn{1}{|c|}{Fault 1}&	\multicolumn{1}{c|}{$10.27$}    &\multicolumn{1}{c|} {$6.45$}    &\multicolumn{1}{c|} {$7.70$}    &\multicolumn{1}{c|} {$8.42$}    \\\hline
\multicolumn{1}{|c|}{Fault 2}&	\multicolumn{1}{c|}{$10.38$}    &\multicolumn{1}{c|} {$6.37$}    &\multicolumn{1}{c|} {$7.63$}    &\multicolumn{1}{c|} {$8.15$}    \\\hline
\multicolumn{1}{|c|}{Fault 3}&	\multicolumn{1}{c|}{$10.33$}    &\multicolumn{1}{c|} {$6.24$}    &\multicolumn{1}{c|} {$7.56$}    &\multicolumn{1}{c|} {$8.50$}    \\\hline
\multicolumn{1}{|c|}{Fault 4}&	\multicolumn{1}{c|}{$10.46$}    &\multicolumn{1}{c|} {$6.53$}    &\multicolumn{1}{c|} {$8.07$}    &\multicolumn{1}{c|} {$9.04$}    \\\hline
\multicolumn{1}{|c|}{Fault 5}&	\multicolumn{1}{c|}{$10.50$}    &\multicolumn{1}{c|} {$5.61$}    &\multicolumn{1}{c|} {$8.01$}    &\multicolumn{1}{c|} {$9.12$}    \\\hline

	\end{tabular}
\end{table}

The results shown in Table \ref{table:1D_smooth} reveal that the values of $SMI$ for $TSI$ are, consistently, the greatest amongst all studied indices for all five fault events, ranging from 10.27 to 10.50. Whereas the values of this metric for $ROMA$ is the smallest amongst the indices for all studied scenarios, ranging within 5.61 and 6.45. The values of $SMI$ for $TPE$ and $TKE$ are stand second and third in smoothness ranking. These results outline that for a 1-dimensional surface, the $TSI$ is the most smooth index and $TPE$, $TKE$, and $ROMA$ follow.

\subsubsection{Two-dimensional Analysis}

{The values of computed $SMI$ for two-dimensional surface corresponding to a change of loads 2 and 3 as the variables are presented in Table \ref{table:2D_smooth}. The results from a change of loads 1 and 3 and loads 1 and 2 as the only the variables in the system with other load kept constant, are similar. }

\begin{table}[h]
	\centering
	\caption{The results from smoothness analysis of stability indices -- two dimensional -- {Change of loads 2 and 3 as the variables while load 1 remains constant}}
	\label{table:2D_smooth}
	\begin{tabular}{ccccc} \hline
		
\multicolumn{1}{|c|} {Index}         &\multicolumn{1}{c|}  {TSI}      & \multicolumn{1}{c|} {ROMA}      &\multicolumn{1}{c|}  {TKE}    &\multicolumn{1}{c|}  {TPE} \\ \hline

\multicolumn{1}{|c|}{Fault 1}&	\multicolumn{1}{c|}{$8.19$}&\multicolumn{1}{c|}{$5.05$}&\multicolumn{1}{c|}{$6.48$}&\multicolumn{1}{c|}{$6.48$} \\\hline
\multicolumn{1}{|c|}{Fault 2}&	\multicolumn{1}{c|}{$8.30$}&\multicolumn{1}{c|}{$4.78$}&\multicolumn{1}{c|}{$6.45$}&\multicolumn{1}{c|}{$6.73$} \\\hline
\multicolumn{1}{|c|}{Fault 3}&	\multicolumn{1}{c|}{$8.24$}&\multicolumn{1}{c|}{$4.78$}&\multicolumn{1}{c|}{$6.38$}&\multicolumn{1}{c|}{$7.01$} \\\hline
\multicolumn{1}{|c|}{Fault 4}&	\multicolumn{1}{c|}{$8.40$}&\multicolumn{1}{c|}{$5.02$}&\multicolumn{1}{c|}{$6.74$}&\multicolumn{1}{c|}{$7.87$} \\\hline
\multicolumn{1}{|c|}{Fault 5}&	\multicolumn{1}{c|}{$8.44$}&\multicolumn{1}{c|}{$4.21$}&\multicolumn{1}{c|}{$6.91$}&\multicolumn{1}{c|}{$7.80$} \\\hline

	\end{tabular}
\end{table}

The results shown in Table \ref{table:1D_smooth} reveal that $TSI$ and $ROMA$ indicate the greatest and smallest values for $SMI$ for all studied fault events, ranging from 8.19 to 8.44 for $TSI$ and 4.21 to 5.05 for $ROMA$. The values of this metric for $TPE$ and $TKE$ range within 6.48 and 7.87 and 6.38 and 6.91, respectively. These results conclude that for a 2-dimensional surface, the $TSI$ is the most smooth index and $TPE$, $TKE$, and $ROMA$ follow. 

\subsubsection{Three-dimensional Analysis}

{The values of computed $SMI$ for smoothness analysis of a three-dimensional surface corresponding to a change of loads 1, 2 and 3 as the variables are presented in Table \ref{table:3D_smooth}.}

\begin{table}[h]
	\centering
	\caption{The results from smoothness analysis of stability indices -- three dimensional  -- {Change of loads 1, 2 and 3 as the variables}}
	\label{table:3D_smooth}
	\begin{tabular}{llllllllllllllllllll} \hline

\multicolumn{1}{|c|} { Index}  & \multicolumn{1}{c|}{TSI} & \multicolumn{1}{c|}{ROMA} & \multicolumn{1}{c|}{TKE} & \multicolumn{1}{c|}{TPE}  \\ \hline

\multicolumn{1}{|c|}{Fault 1}&	\multicolumn{1}{c|}{$7.94$}&\multicolumn{1}{c|}{$5.07$}&\multicolumn{1}{c|}{$6.86$}&\multicolumn{1}{c|}{$6.11$} \\\hline
\multicolumn{1}{|c|}{Fault 2}&	\multicolumn{1}{c|}{$8.02$}&\multicolumn{1}{c|}{$4.77$}&\multicolumn{1}{c|}{$6.82$}&\multicolumn{1}{c|}{$6.52$} \\\hline
\multicolumn{1}{|c|}{Fault 3}&	\multicolumn{1}{c|}{$7.98$}&\multicolumn{1}{c|}{$4.72$}&\multicolumn{1}{c|}{$6.73$}&\multicolumn{1}{c|}{$6.81$} \\\hline
\multicolumn{1}{|c|}{Fault 4}&	\multicolumn{1}{c|}{$8.12$}&\multicolumn{1}{c|}{$5.00$}&\multicolumn{1}{c|}{$7.06$}&\multicolumn{1}{c|}{$6.63$} \\\hline
\multicolumn{1}{|c|}{Fault 5}&	\multicolumn{1}{c|}{$8.16$}&\multicolumn{1}{c|}{$4.34$}&\multicolumn{1}{c|}{$7.23$}&\multicolumn{1}{c|}{$7.82$} \\\hline
		
		\end{tabular}
\end{table}

The results presented in Table \ref{table:3D_smooth} highlight that the values of smoothness metric for $TSI$ and $ROMA$ are consistently the largest and smallest for all studied faults. This suggests that the three-dimensional surfaces created using these two transient stability indices are the most and least smooth surfaces amongst the studied surfaces, respectively. The surfaces constructed by using $TPE$ and $TKE$ indices are second and third in terms of smoothness amongst investigated indices.

{The reason for the $TSI$ to show the greatest level of smoothness can be justified by its simple and linear relationship to the rotor angle difference, as presented by (\ref{eq:TSI}). $ROMA$ is computed using rate of change of acceleration of generators, defined by (\ref{eq:ROCOF}), which varies significantly depending on operational point of each generator prior to fault. $TKE$ is computed using summation of quadratic functions with different weights in which different angular momentum of generators are a factor, given by (\ref{eq:TKE}). Thus, its behavior is non-linear and, therefore, its smoothness is weakened. $TPE$ is computed using integral of dot product of two dynamic variables of pair-generators, their frequency deviation and transient active power, defined by (\ref{eq:TPE}). Therefore, it is a non-linear function and its smoothness in response to change of system's variables is influenced.}

Fig. \ref{Smooth_grpah} visualizes the $SMI$ for studied indices in multiple dimensions. The results shown in this figure reinforce the consistency of this smoothness measure and how suitably it captured the smoothness of indices in an multi-dimensional analysis.}

\begin{figure}[h]
	\centering
	\includegraphics[width=0.445\textwidth]{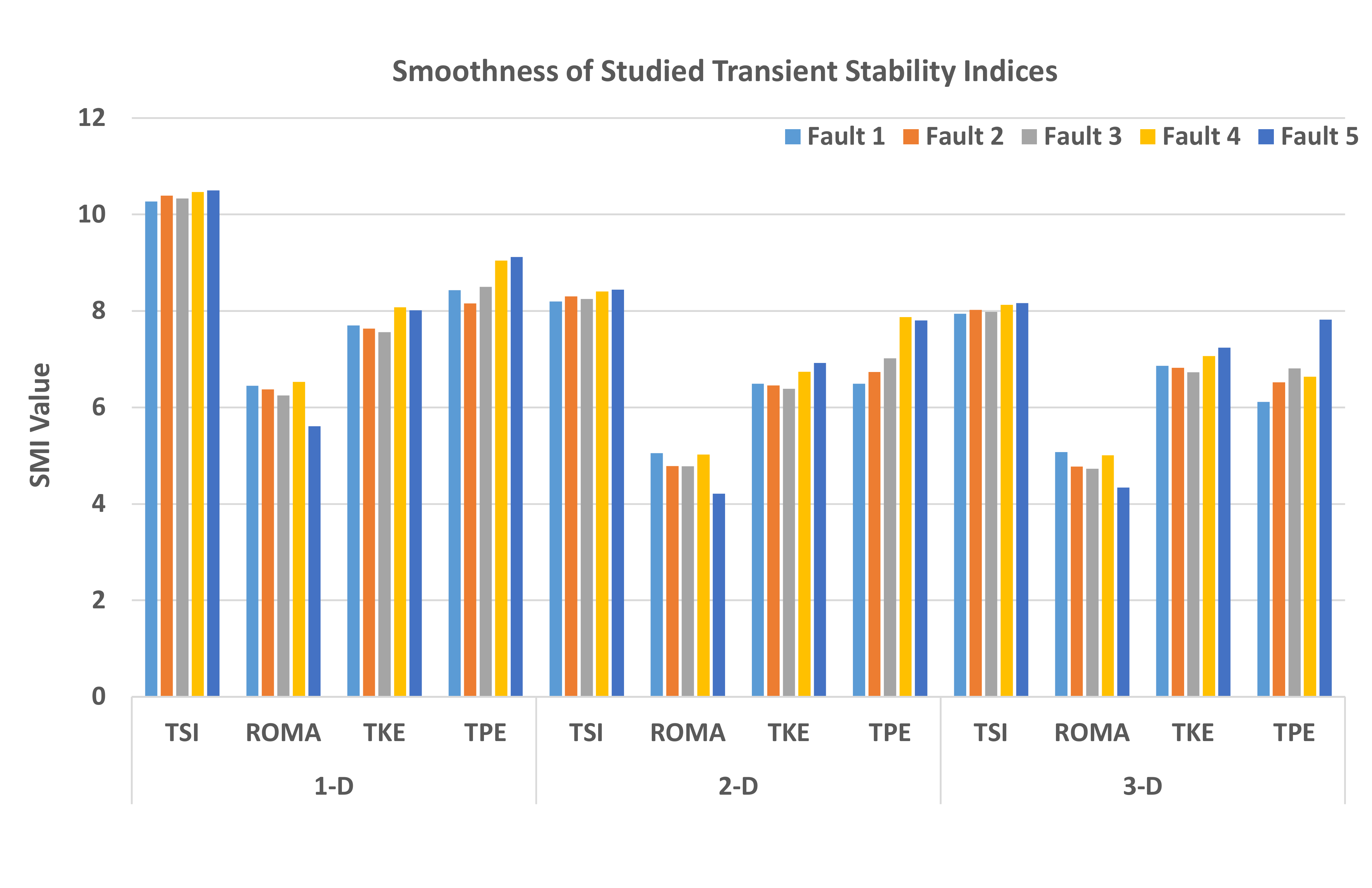}
	\caption{The results from smoothness analysis of studied transient stability indices}
	\label{Smooth_grpah}
\end{figure}

\begin{figure*}[!]
	\centering
	\begin{subfigure}[b]{0.3\textwidth}
		\includegraphics[width=0.999\textwidth]{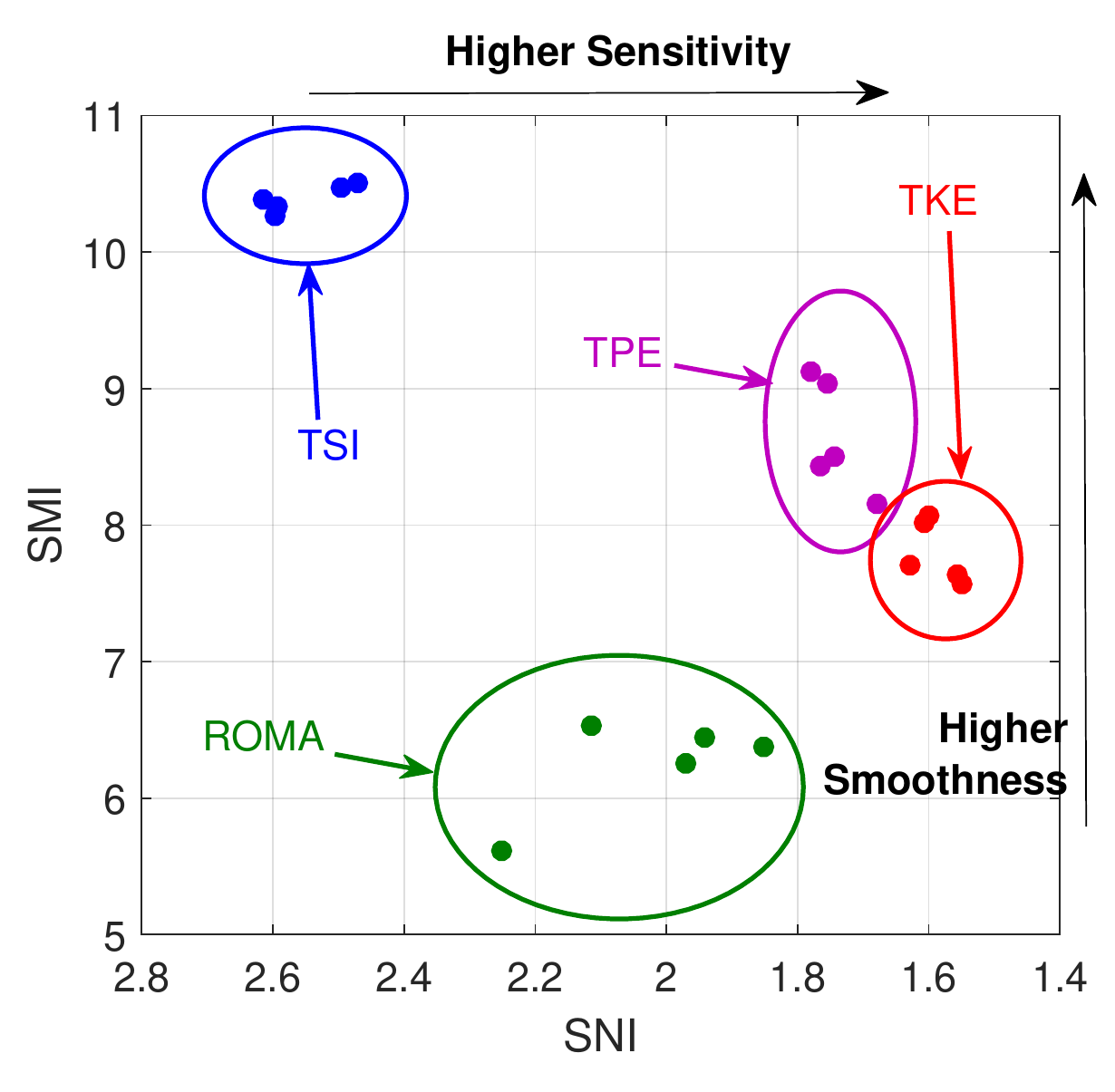}
		\caption{One-dimensional}
		\label{sens_smooth1}
	\end{subfigure}
	\begin{subfigure}[b]{0.3\textwidth}
		\includegraphics[width=0.999\textwidth]{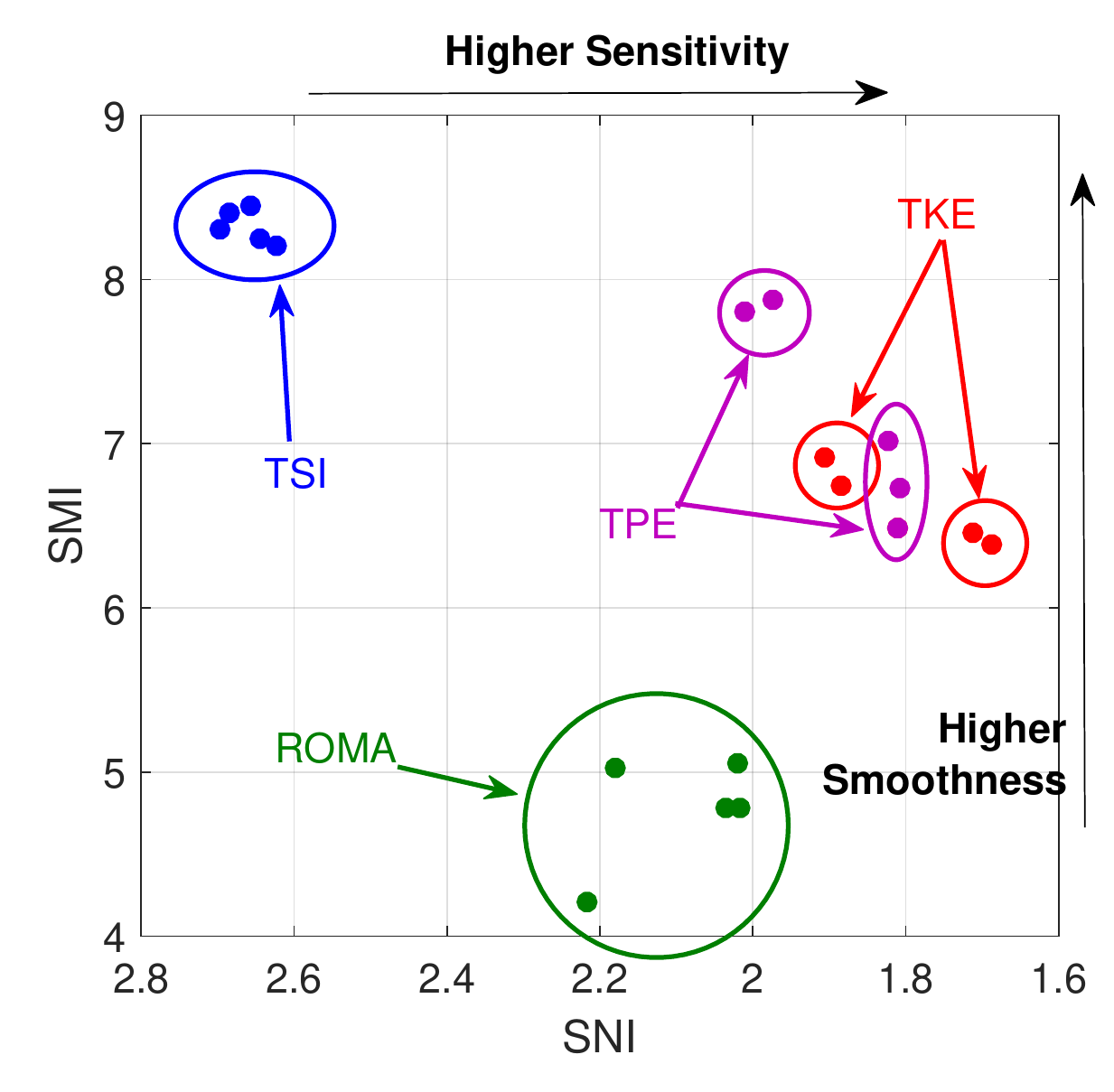}
		\caption{Two-dimensional}
		\label{sens_smooth2}
	\end{subfigure}
	\begin{subfigure}[b]{0.3\textwidth}
		\includegraphics[width=0.999\textwidth]{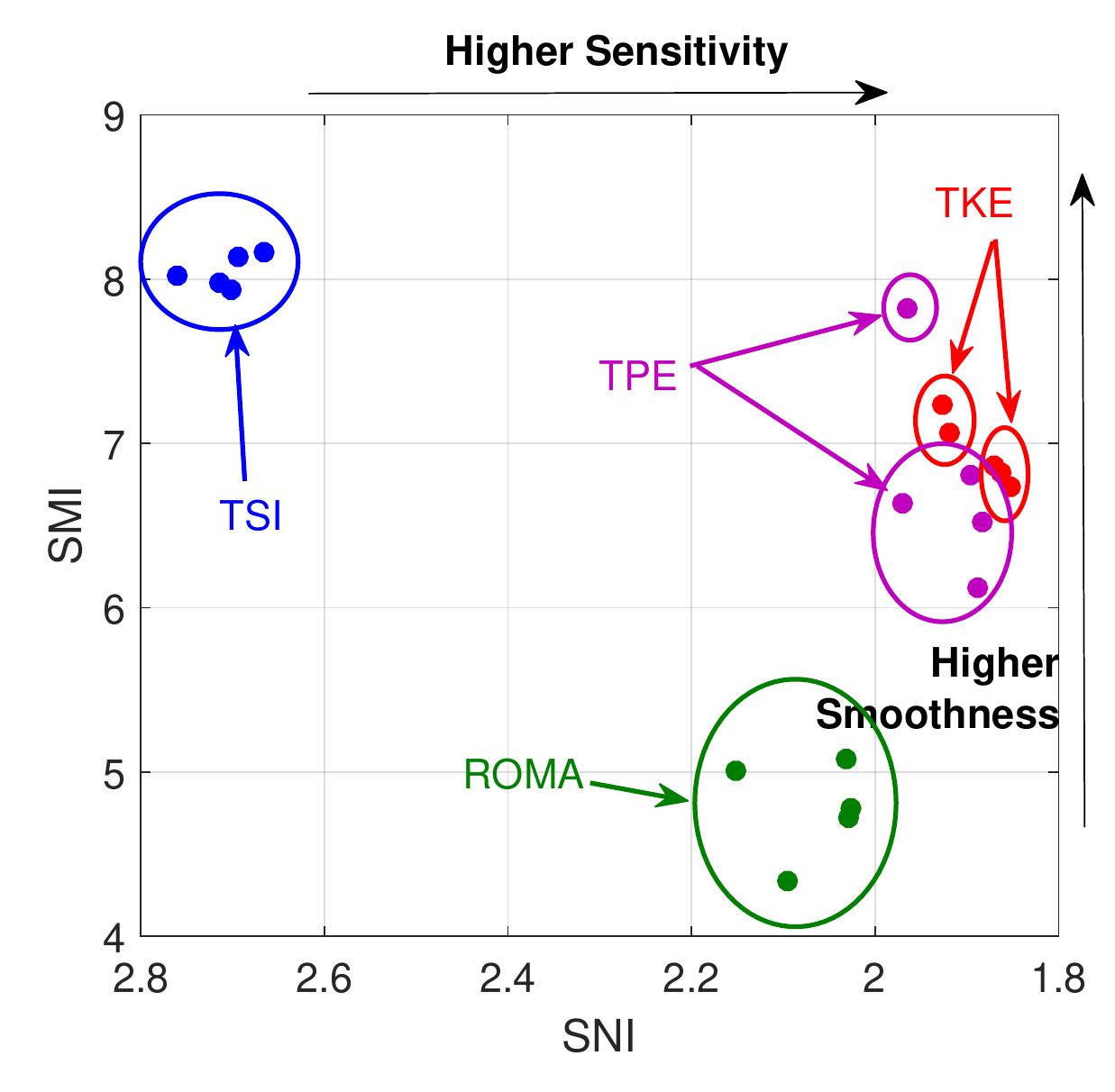}
		\caption{Three-dimensional}
		\label{sens_smooth3}
	\end{subfigure}
	
	\caption{Smoothness vs sensitivity for studied indices}
	\label{sens_smooth}
\end{figure*}

\subsection{Discussion}

In two previous sections, the results from sensitivity analysis and smoothness analysis of four commonly used transient stability indices in power industry, $TSI$, $ROMA$, $TPE$, and $TKE$, are shown. The results included single-dimensional analysis which reflects variation of a single load in the system, two-dimensional analysis which addresses variation of two loads in a system, and three-dimensional analysis which addresses variation of three and all of the loads in this system. 

{By looking at the consensus among all presented results in this paper, a clear consistency across multi-dimensional analysis can be seen. This highlights the suitability of suggested data analysis methods.}

The point to note from the results presented in section \ref{sec:sens}, \textit{Sensitivity Analysis}, is that, regardless of dimensionality of the system, $TKE$ is the most sensitive index for transient stability analysis in power systems. $TPE$, $ROMA$, and $TSI$ also are sensitive to a change of variables in the system, however, by lesser degrees.

The point to note in this study, from the results presented in section \ref{sec:smooth}, \textit{Smoothness Analysis}, is that, regardless of dimensionality of the system, the surface created by the $TSI$ stability index is the smoothest for transient stability analysis in power systems. The surfaces created by the $TPE$, $TKE$, and $ROMA$ are smooth as well, however, to a lower degree. 

The main purpose of this study has been to identify the most suitable transient stability index that can be utilized for different facets of power system stability risk analysis and to conduct further studies regarding operation and control of power systems with a higher degree of complexity and uncertainty. {Ideally, a single index is desired to, simultaneously, offer the highest level of sensitivity and smoothness. However, each of them have their own limitations. Thus, the desired stability index can be scrutinized and identified by a fair trade-off between these two measures.} Fig. \ref{sens_smooth} visualizes the smoothness vs. sensitivity for each of the studied indices in various dimensional space for the different faults considered.

By considering the presented results in various dimensions and presented plot in Fig. \ref{sens_smooth}, it can be concluded that $TPE$ is the most suitable transient stability index for the purpose of further studies. This index offers consistently high levels of both sensitivity and smoothness.

$TSI$ offers the greatest level of smoothness with respect to variability of system's operational condition. However, its major weakness is its lower sensitivity which makes it less attractive. Similarly, $TKE$ shows the greatest level of sensitivity while its lower smoothness is a disadvantage for this index. Finally, $ROMA$ features low sensitivity and low smoothness and is the least valuable stability index in this sense.

\section{Conclusion}


This work attempted to identify the stability indicators that can be used for different facets of future power system stability risk analysis with higher dimensionality and complexity. To evaluate the suitability of the desired index, its sensitivity to a change of variable and operational conditions of system as well as smoothness of the surface created by the given index in a multiple-dimensional space were investigated. 

This research provided a comparison among the transient stability indices established in literature. These indices included rotor-angle difference based transient stability index (TSI), rate of machine acceleration (ROMA), transient kinetic energy (TKE), and transient potential energy (TPE). A 3-machine, 9-bus standard test system was used as a case study. 

The results suggest that $TPE$ is the most suitable transient stability index for the purpose of further studies as it offers consistently high levels of both sensitivity and smoothness.

Further investigation will include developing mathematical framework to efficiently identify operating conditions or system contingencies that will lead to instability in a high-dimensioned search-space using the identified suitable indices in larger power systems such as NETS-NYPS and Great Britain networks. It will focus on development of an efficient method to find critically unstable system conditions of the system, including low probability high impact events, to ensure that sufficient samples are used and various facets of system operation are included.

\section*{Acknowledgment}

The authors would like to thank Research Councils UK for financial support of this research through the HubNet consortium (grant number: EP/N030028/1).

\bibliography{IREP}
\bibliographystyle{IEEEtran}

\end{document}